\documentclass[%
 reprint,
showpacs,preprintnumbers,
 amsmath,amssymb,
 aps,
]{revtex4-2}

\usepackage{graphicx}
\usepackage{dcolumn}
\usepackage{bm}

\begin{document}


\title{Broken-SUSY dynamic reflectionless optical amplifier}
\author{Ugo Tricoli}
\email{ugo.tricoli@onera.fr}
\author{Leon Schlemmer}%

\affiliation{%
 ONERA, The French Aerospace Lab, Base A\'erienne 701,13661 Salon Cedex AIR, France
}%

\begin{abstract}
	
	    Broken-supersymmetry is used to define a reflectionless active cavity capable of amplifying electromagnetic radiation in the visible. The approach is analytical through the use of the Darboux transform for the generation of the optical potential and the calculation of its solution, while the transmission/reflection spectra evaluation is done with the Transfer Matrix method. Interestingly, the proposed device behaves as a dynamic optical filter amplifying radiation arriving at large angles while for other directions is almost completely transparent. Thus, simply by rotation different functionalities can be obtained. In addition the active filter is reflectionless for all wavelengths and angles of incidence. 
	    The necessary Kramers-Kronig relations are also satisfied
	    being the model compatible with the  Drude model at high frequency i.e. far from absorption resonances. 
	

\end{abstract}

\maketitle


\section{\label{sec:level1}Introduction}

The design of metamaterials with exceptional optical properties has attracted major attention in the last decades due to the possibility of experimental realization offered by recent technological advances \cite{cai2010optical,liu2011metamaterials}. A particularly interesting class of these materials is characterized by supersymmetry (SUSY).

Initially, it has been proposed to use supersymmetric transformations \cite{cooper1995supersymmetry} to generate PT-symmetric potentials, however in the context of quantum mechanics making their experimental realization difficult \cite{cannata1998schrodinger,sinha2002isospectral,salinas2003confluent}.
On the other hand, in optics in the context of optical waveguides, it has been recently realized that supersymmetry can be used in order to restore PT-symmetry or to produce classes of optically equivalent potentials (in terms of reflection and transmission amplitudes) with extremely different spatial distribution of optical properties \cite{miri2013supersymmetry,miri2013supersymmetric,miri2014susy}. SUSY has also been used to generate discrete real optical potentials that are reflectionless \cite{longhi2015supersymmetric} or PT-symmetric fiber optical lattices \cite{regensburger2013observation,longhi2014bound,longhi2014optical}. 
What makes SUSY very attractive for the design of new optical devices is the possibility to define different spatial refractive index distributions (superpartners) having the same scattering spectra (angularly and spectrally). For this work we explore the possibilities offered by the generation of superpartners of vacuum, being reflectionless and possessing unit transmission by definition.
%
It is well known that reflectionless optical potentials can be obtained through supersymmetric transformations \cite{longhi2015supersymmetric} with the limitation that exact reflectionless configurations are obtained in practice (with spatial refractive index distributions) only for discrete frequencies. 
Interestingly, we show that the Darboux transform (DT), a type of supersymmetric transformation, introduces (indirectly) dispersion analytically in media that are spatially heterogeneous along one direction, thus allowing to define materials that are reflectionless for a continuum of frequencies. The critical point consists in the fact that the Darboux transform acts on the optical potential which is by definition frequency dependent (i.e. proportional to the square of the incoming wavenumber). 
It is hard to treat dispersion through the DT procedure, so a much easier strategy is to generate potentials for a fixed wavelength (as it is done in SUSY waveguides \cite{miri2013supersymmetry,longhi2015supersymmetric,miri2013supersymmetric}). As a result the generated potential is frequency independent and
consequently when converting it back to the optical properties (e.g. the refractive index), these become dispersive. This scheme is very different from traditional methods in metamaterials where the mathematical transformations are applied directly on the optical properties so that the potential is by definition frequency dependent (with the same dependence as electromagnetic waves). On the contrary, in the Darboux scheme the potential 
is calculated at a single frequency and wave dispersion must be reintroduced for example by using the optical potential definition, thus passing the electromagnetic wave dispersion on the optical properties.
An important remark should be done here, the electromagnetic wave dispersion is normally separated (factorized out in the optical potential definition) from the material dispersion coded in its optical properties. In this utilization of the DT, the electromagnetic wave dispersion is imposed on the optical properties of the DT defined material. Consequently, the application of a DT implies also that a particular material is chosen (according to the frequency dispersion of e.g. the refractive index). 
This has a heavy consequence on the Darboux generated optical properties i.e. both the real and the imaginary parts of the refractive index share the same dispersion. This sort of artificial electromagnetic wave dispersion imposed on the refractive index is compatible with physical materials (as represented by the Drude model \cite{fox2001optical}) only in the limit of high frequencies i.e. far from absorption resonances thus for very small absorption. 
Indeed, in order to satisfy the necessary Kramers-Kronig relations \cite{altarelli1972superconvergence}, the imaginary part of the refractive index should be kept sufficiently small (few orders of magnitude smaller than the real part). 
To summarize, our study is directed towards the calculation of transmission and reflection spectra in order to make use of SUSY to design optimal optical filters. In this framework we take into account spatial distributions of the refractive index possessing resonances that can be calculated as bound states of a Schr{\"o}dinger-like equation (transversal modes). 
Interestingly, when SUSY is broken (for complex energy states) anomalous transmission is observed i.e. the transmission coefficient becomes larger than one \cite{mostafazadeh2009spectral,mostafazadeh2015physics}. This is a manifestation of the fact that the system is open towards the external environment i.e. external energy is used to generate anomalous transmission. 
This behavior is typical of open systems \cite{persson2000observation} and its realization in optics is associated with many extraordinary effects \cite{miri2014susy,lin2011unidirectional,ruter2010observation,makris2011mathcal,regensburger2012parity,heinrich2014observation,longhi2018parity,el2018non}. Hence, classical optical systems with broken SUSY provide the unique possibility of exploiting the special features normally associated to quantum open systems. 
Thus, a broken SUSY active optical cavity is proposed which can be used as a dynamic optical amplifier. Indeed, for large incidence angles amplification is obtained, while for small angles the device is almost transparent (due to SUSY breaking). Notably, being the superpartner of vacuum, the device is reflectionless for all wavelengths and angles. Interestingly, its dynamic properties can be accessed by rotating the device with respect to the incoming radiation direction.

\section{\label{sec:level2}Scattering theory }

We consider scattering of a monochromatic electromagnetic wave by an inhomogeneous dielectric slab of finite thickness along $\hat{x}$ but infinite extension in the $yz$-plane with a varying refractive index in the $\hat{x}$ direction $n(x)$. The slab is surrounded by a non-absorbing and non-dispersive background medium with refractive index $n_{b}$. Taking into account only a scalar electric field (i.e. the component $E_{z}$ corresponding to a TE-polarized wave) with the time dependence factorized as $\exp({-i\omega t})$, the curl Maxwell equations can be rewritten as the Helmholtz wave equation
\begin{equation}
	\label{Helmolholtz}
	\frac{\partial^2 E_{z} }{\partial x^2} + \frac{\partial^2 E_{z} }{\partial y^2} + [k_{0}n(x)]^2  E_{z} = 0,
\end{equation}
with $k_{0}=\omega/c$ the incoming wavenumber in vacuum and $n(x)= n_b + \Delta n(x)$ the spatial dependent refractive index, with the last term being the optical contrast which is complex valued and vanishes at infinity. For a general angle of incidence the solution to Eq.(\ref{Helmolholtz}) can be factorized through an amplitude and a phase as
$
\label{sol1}
E_z(x,y) = \psi(x) \exp(ik_y y)
$,
with the incoming wavevectors related by $k_y^2= [k_{0}n_b]^2-k_x^2$. Substituting into Eq.(\ref{Helmolholtz}) and  upon addition/subtraction of the term $[k_{0}n_b]^2 \psi(x)$ we get the equation for the amplitude $\psi(x)$
\begin{equation}
	\label{Schroedinger}
	[-\frac{d^2 }{d X^2} + V(X)] \psi(X) = e \psi(X),
\end{equation}
which is in a Schr{\"o}dinger-like form $H\psi=e\psi$, changing the variable $X=k_0 x$ and defining the optical potential as $V(X)=- [n(X)^2-n_b^2]$, the Hamiltonian $H=-d^2 \psi(X) /d X^2 + V(X)$ and the energy $e=k_x^2/k_0^2=[(k_{0}n_b)^2 - k_y^2]/k_0^2$, i.e. the square of the component of the wavevector of the incoming wave along the direction of variation of the refractive index.
%
%
Due to the reduction of the optical Helmholtz wave equation to a Schr{\"o}dinger-like equation, it is possible to apply in optics techniques originally developed in quantum mechanics (e.g. SUSY \cite{cooper1995supersymmetry,cannata1998schrodinger}).

\section{\label{sec:level3}Supersymmetry in Optics}

Supersymmetry is usually employed to engineer bound states in optical structures serving as waveguides \cite{miri2013supersymmetry,longhi2015supersymmetric,miri2013supersymmetric}, hence for grazing incidence only. Nevertheless, for incidence not limited to directions near to the principal optical axis of the structure, supersymmetry can be exploited in optics to design completely different structures that are indistinguishable from outside (as it was noted in \cite{miri2013supersymmetric,miri2014susy}).
For the purpose of this work, the most attractive feature of supersymmetric transformations lies in the possibility to relate two different optical potentials sharing the same transmission and reflection spectra. 
Here we make use of a particular realization of supersymmetry i.e. the DT \cite{darboux1882proposition,cannata1998schrodinger} which also allows building solvable profiles of the refractive index \cite{krapez2017sequences}.

In general, given a fixed incoming wavenumber $k_{sc}=\omega_{sc} / c$ (which serves as a spatial scaling factor for the DT i.e. $X = k_{sc} x$ as in \cite{tricoli2020susy}), the main action of a DT is to relate the spectral properties of a pair of Schr{\"o}dinger Hamiltonians through 
a superpotential $\sigma$ defined as the
logarithmic derivative of a solution of the Schr{\"o}dinger equation $\psi(X,\epsilon)$ for a given potential $V(X)$ for some value of the transformation energy $\epsilon$ (in units of $k^2_{sc}$) i.e. $\sigma(X,\epsilon)=\frac{1}{\psi(X,\epsilon)}\frac{d\psi(X,\epsilon)}{dX}$. 
The partner potential of $V(X)$ is found through the optical DT,
\begin{equation}
\label{DT}
V_{DT}(X) = -V(X)+ 2\sigma(X,\epsilon)^2 + 2\epsilon,
\end{equation}
which is in units of $k^2_{sc}$ too, and the corresponding Schr{\"o}dinger equation has the solution (for a general $k_{0}$):
\begin{equation}
	\label{DT_sol}
	\psi_{DT}(X,k_{0}) = [-\frac{d}{dX}+ \sigma(X,\epsilon)]\psi(X,k_{0}). 
\end{equation}
%
Generally speaking, the DT can generate two types of partner potentials that are referred to as broken and unbroken SUSY partners. When the DT is adding a bound state to the spectrum of the starting Hamiltonian, and this new state it is found to be the new ground state of the transformed Hamiltonian, SUSY is preserved (unbroken). The two potentials are indistinguishable from outside i.e. they have the same reflection and transmission spectra (as it was already noted in \cite{miri2013supersymmetric,miri2014susy} at a fixed frequency for all angles of incidence). On the contrary, if the DT is adding a bound state which is not the new ground state of the transformed Hamiltonian, SUSY is broken. In this case the two super partners are distinguishable from outside even if their spectra still share some properties.
%
%
This can be exploited straightforwardly to generate analytically reflectionless optical potentials. Indeed, by transforming the null potential ($V(X)=0$) which has no reflection and unit transmission by definition, we can define its supersymmetric partner being reflectionless for all incoming frequencies (and angles of incidence) by construction (thus generalizing SUSY applications as in \cite{miri2013supersymmetric} that were just for all angles at a fixed frequency). Interestingly its broken-SUSY partner is still reflectionless but the transmission can differ from unity depending on the sign of the imaginary part of the transformation energy $\epsilon$ (see Section \ref{sec:level3} for examples).
Nevertheless, the Darboux transform introduces a dispersion on the optical properties when the potential is converted to the refractive index distribution i.e.
\begin{equation}
\label{index}
n_{DT}(X,\omega) = \sqrt{1-V_{DT}(X)k_{sc}^2/k_{0}^2} 
\end{equation}
by using the optical potential definition $V(X)=-k_{0}^2(n^2(X)-n^2_b)$ and considering air as the background medium i.e. $n_b=1$.
This can be seen by looking at the solution of the transformed Schr{\"o}dinger equation expressed in Eq.(\ref{DT_sol}), which depends on $k_{0}=\omega/c$.
Thanks to this (electromagnetic wave) dispersion, the Darboux generated refractive index profile is analytically defined as a superpartner for every incoming wavelengths and directions.
An important remark is needed here. If dispersion is neglected, 
SUSY can be completely broken thus making the two potentials non-being superpartner anymore for all the spectrum (they remain superpartner only for the unique wavelength which is chosen to perform the DT, and some integer multiples of it \cite{longhi2015supersymmetric}). Moreover, if the dispersion of Eq.(\ref{index}) is modified, similarly the two partner potentials are no longer superpartners and the analyticity of all the procedure is lost. 

Finally, as it was already discussed in Section \ref{sec:level1}, the dispersive index spatial profiles as generated by a single DT are physical in the sense that they are compatible with the Kramers-Kronig frequency relations only in the high frequency limit i.e. far from resonances where the Drude model is valid and absorption is very small.

\section{\label{sec:level3}Results}

\begin{figure}[h]
	\centering
	\includegraphics[width=4cm]{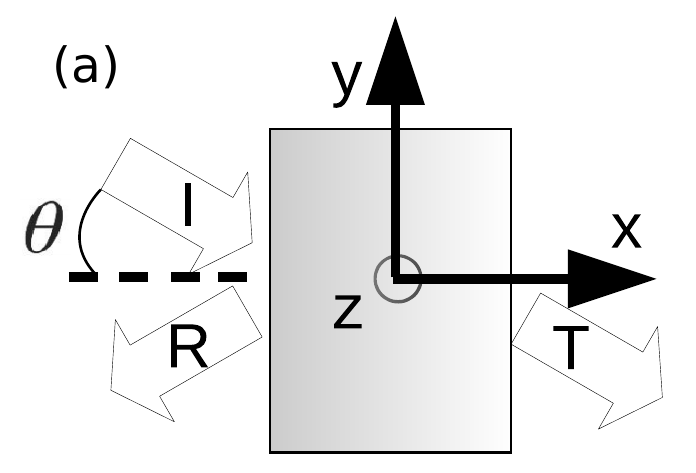}
	\includegraphics[width=4cm]{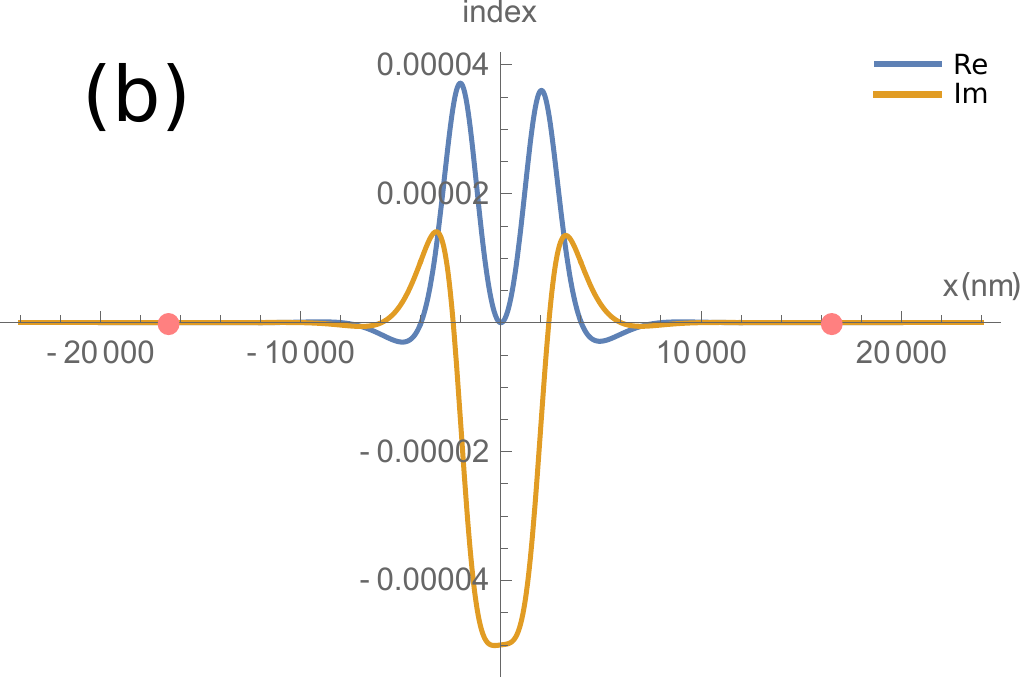}	
	\includegraphics[width=4cm]{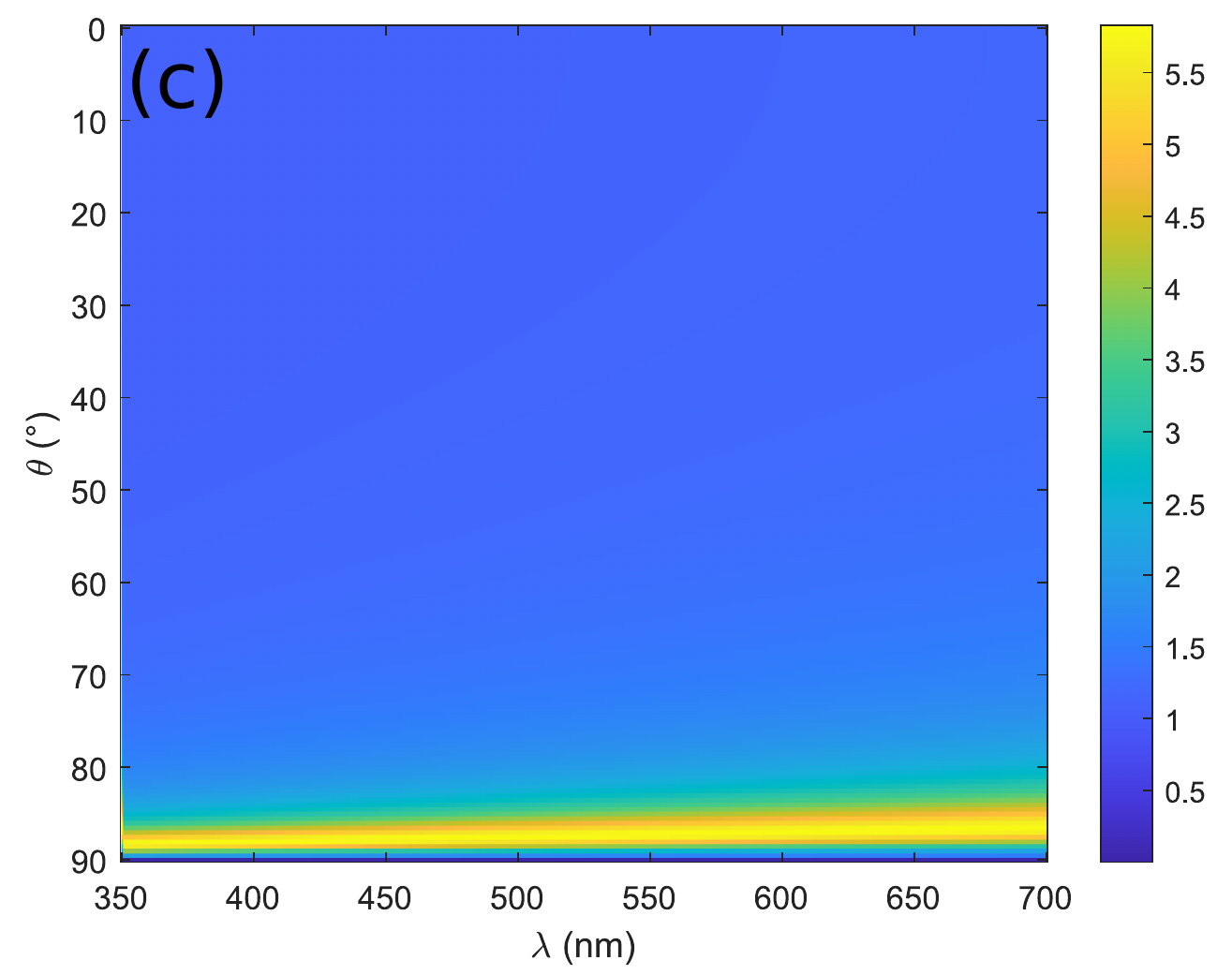}
	\includegraphics[width=4cm]{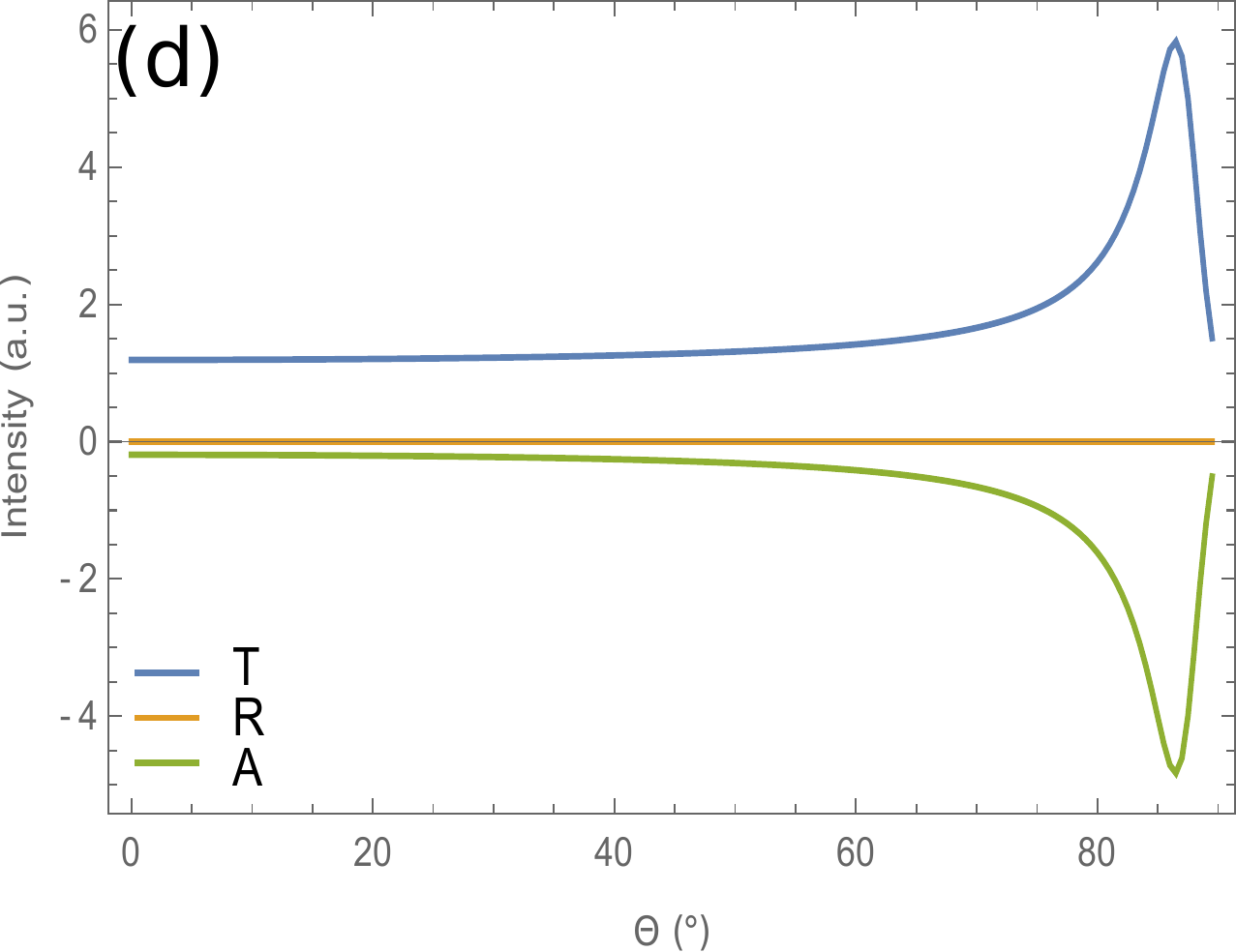}	
	\caption{The reflectionless active cavity for amplification: (a) the scattering setup (b) the refractive index variation $\Delta n(x)$ at $k_{sc}$ (c) the spectro-angular spectrum in the visible (d) the angular spectrum at $\lambda=700$ nm. All the results are for $k_{sc}=2 \pi/80 \rm{nm}^{-1}$. The red dots indicate the extremes for the Transfer Matrix calculation for the analytical evaluation of transmission and reflection coefficients.
	}
	\label{Fig1}
\end{figure}

Given some $\epsilon$, we consider the supersymmetric partner of the null potential $V(X)=0$  which can be obtained through a single DT of $\psi(X,\epsilon)= C_1 \cos(\sqrt{\epsilon}X) + C_2 \sin(\sqrt{\epsilon}X)$, giving 

\begin{equation} 
\label{V2}
V_{DT}(X) = 2 \epsilon \left[1 +  \left( \frac{ C_2 - C_1 \tan (\sqrt{\epsilon}X)  }{ C_1 + C_2 \tan (\sqrt{\epsilon}X)  } \right) ^2 \right]
\end{equation}
where for the present study we took $\epsilon=0.00005 i$ and $C_1 = 1, C_2=0.01$. 
Firstly, we note that the obtained index distribution (see  Fig.(\ref{Fig1})) has very small imaginary part as it is required by the Kramers-Kronig relations. In addition, the profile has no spatial symmetry and has a typical width of few micrometers. To evaluate its spectral response, the analytical Transfer Matrix method is used as explained in \cite{mostafazadeh2020transfer} using the analytical solution for the field as expressed by Eq.(\ref{DT_sol}).
Considering the angular spectrum, the device is almost transparent for incident angles lower than $70^\circ$ (with transmission slightly larger than unity). For incoming directions around $85^\circ$, the device behaves as an amplifier for the visible spectrum. As already mentioned, the proposed filter remains reflectionless for all angles and wavelengths (in Fig.(\ref{Fig1}) it is just shown for a single wavelength and all angles). Thus, the device can be considered as a reflectionless active cavity capable of amplifying radiation impinging at large angles. Interestingly, if amplification is not required, simply by rotating the device, it is possible to bypass it using its transparency in the rest of the angular spectrum.

\section{\label{sec:level4}Conclusions}

We demonstrate how using supersymmetry we can define refractive index spatial distributions that can be used as a dynamical optical amplifier. In particular, the proposed filter is reflectionless by construction and is amplifying the incoming radiation for large angles close to $85^\circ$ while it is almost transparent for smaller angles. The device can be rotated to achieve dynamism i.e., by changing the incoming direction, the same filter can be used as an amplifier or it can be bypassed exploiting its transparency for smaller angles. This is valid in transmission while reflection is always completely suppressed for all wavelengths and angles due to the fact that vacuum is transformed through a DT. This result is also related to the conversion of the optical potential into the refractive index i.e. the Darboux transform introduces electromagnetic wave dispersion. The effect of the introduced dispersion is to change the amplitude of the spatial distribution of the refractive index accordingly to the incoming wavelengths, allowing no reflection for all the spectrum. Luckily for physical feasibility, this dispersion corresponding to the Drude model at high frequencies, is compatible with the Kramers-Kronig relations for frequencies sufficiently far from absorption resonances. A good candidate material to realize the proposed device for a portion of the visible spectrum (the one with the right dispersion) would be $SiO_2$, also suitable for use in direct laser writing \cite{ocier2020direct,porte2021direct}, which appears as the most directly appropriate technique to realize gradient index spatial distributions. However, a challenge remains which is linked to the possibility to control absorption and gain in space which can be achieved through doping.

\bibliography{SUSY_Rless}

\end{document}